\documentclass[12pt]{article}
\usepackage{lmodern} 
\usepackage[english]{babel}
\usepackage{amsmath}
\usepackage{amsfonts}
\usepackage{amssymb}
\usepackage{graphicx}

\usepackage[pdftex]{hyperref}
\hypersetup{
    plainpages=false,       
    unicode=false,          
    pdftoolbar=true,        
    pdfmenubar=true,        
    pdffitwindow=false,     
    pdfstartview={FitH},    
    pdfauthor={Andrew W.L. Fleck},    
    pdfnewwindow=true,      
    colorlinks=false,        
	linkbordercolor={1 0 0}, 	 
	citebordercolor={0 1 0}, 	
	urlbordercolor ={0 1 1},    
    linkcolor=blue,         
    citecolor=green,        
    filecolor=magenta,      
    urlcolor=cyan           
}

\usepackage{bm} 

\usepackage[affil-it]{authblk} 

\usepackage{verbatim}

\usepackage{color,soul}

\usepackage{tikz}
\usetikzlibrary{calc,decorations.markings}

\usepackage{caption}
\usepackage{subfig}
\usepackage{wrapfig,lipsum,booktabs}

\usepackage{epigraph}

\usepackage{sectsty}
\sectionfont{\centering}

\numberwithin{equation}{section}
\numberwithin{figure}{section}
\numberwithin{table}{section}

\usepackage{amsthm}
\newtheorem{theorem}{Theorem}[section]

\newtheorem{lemma}[theorem]{Lemma}
\newtheorem{definition}{Definition}[section]

\newcommand{\stablefun}[4][\infty]{ \int_{0}^{#1} e^{- \sigma_{#4}^\alpha t^\alpha } {#2} \cos( {#3} t - a \beta_{#4} \sigma_{#4}^\alpha t^\alpha) dt }

\newcommand{\inprod}[2]{ {#1}^{T}{#2} }

\newcommand{\expect}{\mathbb{E}}
\newcommand{\Cexpect}[2]{ \mathbb{E}[{#1}|{#2}]}

\usepackage{mathbbol} 
\newcommand{\var}{\mathbb{Var}}

\newcommand{\cov}{\mathbb{Cov}}

\newcommand{\mel}[2]{ \mathcal{M}\left\lbrace {#2} \right\rbrace \left({#1}\right) }

\newcommand{\lap}[2]{ \mathcal{L}\left\lbrace {#2} \right\rbrace \left({#1}\right) }

\newcommand{\fou}[2]{ \mathcal{F}\left\lbrace {#2} \right\rbrace \left({#1}\right) }

\newcommand{\G}[1]{ \Gamma \left( {#1} \right) }

\newcommand{\sign} {{\mathrm{sign} \,}}

\usepackage[automake,toc,abbreviations]{glossaries-extra}

\newabbreviation{mle}{MLE}{Maximum Likelihood Estimate}

\newabbreviation{gmm}{GMM}{Generalized Method of Moments}

\newabbreviation{cgmm}{CGMM}{Continuous Generalized Method of Moments}

\newabbreviation{pdf_words}{PDF}{Probability Density Function}

\newabbreviation{cdf_words}{CDF}{Cumulative Distribution Function}

\newabbreviation{mgf_words}{MGF}{Moment Generating Function}

\newabbreviation{cf_words}{CF}{Characteristic Function}

\newabbreviation{clt}{CLT}{Central Limit Theorem}

\newabbreviation{gclt}{GCLT}{Generalized Central Limit Theorem}

\newabbreviation{edm}{EDM}{Exponential Dispersion Model}

\newabbreviation{cr}{CR}{Cramér–Rao}

\newabbreviation{VaR}{VaR}{Value at Risk}

\newabbreviation{ES}{ES}{Expected Shortfall}

\newabbreviation{TCE}{TCE}{Tail Conditional Expectation}

\newabbreviation{dy}{DY}{Development Years}

\newabbreviation{ay}{AY}{Accident Years}

\newabbreviation{cl}{CL}{Chain Ladder}

\newabbreviation{pcps}{PCPs}{Premium Calculation Principles}

\newabbreviation{capm}{CAPM}{Capital Asset Pricing Model}

\newabbreviation{apt}{APT}{Arbitrage Pricing Theory}

\newabbreviation{wipm}{WIPM}{Weighted Insurance Pricing Model}

\newabbreviation{abrm}{ABRM}{Additive Background Risk Model}

\newglossary*{symbols}{List of Symbols}

\newglossaryentry{Capital_Letter}
{
name={$X$},
sort={Prob},
type=symbols,
description={Random Variables: In general, we denote random quantities by capital letters and deterministic ones by lowercase. }
}

\newglossaryentry{Stoch_Process}
{
name={$\sigma_t$,$r_T$ or $W_t$ etc},
sort={Prob},
type=symbols,
description={Stochastic Processes: We denote families of random variables indexed by time  ($t$ or $T$) as either lower- or uppercase letters with time as the subscript.}
}

\newglossaryentry{pdf}
{
name={$f_{X}(x|\theta)$},
sort={Prob},
type=symbols,
description={Density Functions: The pdf of $X$ at a specific value $x$ given parameters $\theta$ }
}

\newglossaryentry{CDF}
{
name={$F_{X}(x)$},
sort={Prob},
type=symbols,
description={Distribution Functions: The cdf of $X$ at a specific value $x$ i.e. $P(X \leq x)$ }
}

\newglossaryentry{Expectation}
{
name={$\expect[ X ]$},
sort={Prob},
type=symbols,
description={Expected value of a random variable}
}

\newglossaryentry{Variance}
{
name={$\var[ X ]$},
sort={Prob},
type=symbols,
description={Variance of a random variable}
}

\newglossaryentry{Covariance}
{
name={$\cov[ X, Y ]$},
sort={Prob},
type=symbols,
description={Covariance of random variables $X$ and $Y$}
}

\newglossaryentry{mgf}
{
name={$M_X(\tau) = \expect[ e^{\tau X} ]$},
sort={Prob},
type=symbols,
description={Moment generating function of a random variable $X$}
}

\newglossaryentry{cf}
{
name={$\phi_X(\tau) = \expect[ e^{i \tau X} ]$},
sort={Prob},
type=symbols,
description={Characteristic function of a random variable $X$}
}

\newglossaryentry{cgf}
{
name={$K(\tau) = \log( M_X(\tau) ) $},
sort={Prob},
type=symbols,
description={Cumulant generating function of a random variable $X$}
}

\newglossaryentry{cumulant}
{
name={$\kappa_n[X] = \frac{d}{d \tau} K_{X}(0) $},
sort={Prob},
type=symbols,
description={$n$th cumulant of a random variable $X$}
}

\newglossaryentry{outprod}
{
name={$\circ$},
sort={LinA},
type=symbols,
description={Outer product: Given two vectors $\mathbf{x}$ and $\mathbf{y}$  of appropriate length, we have $\mathbf{x} \circ \mathbf{y} = \mathbf{x}  \mathbf{y}^\top $.}
}

\newglossaryentry{vector}
{
name={$\mathbf{X}$ or $\mathbf{x}$},
sort={LinA},
type=symbols,
description={Vectors: Random or deterministic vectors will be identified as boldface quantities. See Random Variables entry for more detail.}
}

\newglossaryentry{matrix}
{
name={$\mathbf{M}_{n \times m}$},
sort={LinA},
type=symbols,
description={Boldface quantities (with subscripts representing dimensions) are reserved for matrices and operators generally. For vectors, that is matrices with one column, we drop the dimension subscripts.}
}

\newglossaryentry{f_op}
{
name={For a function $f(x)$: $\mathcal{A}f$ or $(\mathcal{A}f)(x)$},
sort={LinA},
type=symbols,
description={Cursive capitals are reserved for integral operators.}
}

\newglossaryentry{transpose}
{
name={$\mathbf{M}_{n \times m}^\top $},
sort={LinA},
type=symbols,
description={Matrix Transpose}
}

\newglossaryentry{mellin}
{
name={$\mel{s}{f(x)}$},
sort={IntTrans},
type=symbols,
description={Mellin transform of $f$ given by $\int_0^{\infty} x^{s-1} f(x) dx $}
}

\newglossaryentry{laplace}
{
name={$\lap{s}{f(x)}$},
sort={IntTrans},
type=symbols,
description={Laplace transform of $f$ given by $\int_0^{\infty} e^{-sx} f(x) dx $}
}

\newglossaryentry{fourier}
{
name={$\fou{s}{f(x)}$},
sort={IntTrans},
type=symbols,
description={Fourier transform of $f$ given by $\int_{-\infty}^{\infty} e^{-isx} f(x) dx $}
}

\newglossaryentry{premium}
{
name={$\pi[X]$},
sort={ActSci},
type=symbols,
description={Premium Principle: The premium paid (in currency) for a risk represented by a random variable $X$ }
}

\newglossaryentry{risk_measure}
{
name={$H[X]$},
sort={ActSci},
type=symbols,
description={Risk Measure: A functional over a suitable set of loss random variables mapping losses to \textit{risk capital}  }
}

\glsdisablehyper 
\makeglossaries

\newbool{is_paper}
\booltrue{is_paper} 

\begin{document}

\title{\textbf{Risk Aggregation and Allocation in the Presence of Systematic Risk via Stable Laws}}

\author[a,c]{Andrew W.L. Fleck%
  \thanks{Email: \texttt{afleck@yorku.ca};}}
\author[a]{Edward Furman}
\author[b]{Yang Shen}
\affil[a]{\rm \small \textit{Department of Mathematics and Statistics, York University, Toronto, ON M3J 1P3, Canada}}
\affil[b]{\rm \small \textit{School of Risk and Actuarial Studies, UNSW Sydney, NSW 2052, Australia}}
\affil[c]{BP Trading and Shipping (NA Power)}
\date{}

\maketitle

\begin{quote}

\begin{center}
 \textbf{Abstract}
 \end{center}  
 
In order to properly manage risk, practitioners must understand the aggregate risks they are exposed to. Additionally, to properly price policies and calculate bonuses the relative riskiness of individual business units must be well understood. Certainly, Insurers and Financiers are interested in the properties of the sums of the risks they are exposed to and the dependence of risks therein. Realistic risk models however must account for a variety of phenomena: ill-defined moments, lack of elliptical dependence structures, excess kurtosis and highly heterogeneous marginals. Equally important is the concern over industry-wide systematic risks that can affect multiple business lines at once. Many techniques of varying sophistication have been developed with all or some of these problems in mind. We propose a modification to the classical individual risk model that allows us to model company-wide losses via the class of Multivariate Stable Distributions. Stable Distributions incorporate many of the unpleasant features required for a realistic risk model while maintaining tractable aggregation and dependence results. We additionally compute the Tail Conditional Expectation of aggregate risks within the model and the corresponding
allocations.


\end{quote}

\newpage

\section{Introduction}\label{sec: p1 intro}

Let $\chi$ be the set of random variables that represent the random liabilities of insurance contracts. Call the elements $X$ of $\chi$ \textit{risks}. Mathematically, $X\in \chi$ is a function on some probability space, measurable with respect to the sigma measure. \textit{Risk measures} are then functionals used to assign finite real values or infinite values (corresponding to the \textit{risk capital}) to elements of $\chi$:  

\begin{center}
$H[X]: \chi \rightarrow \mathbb{R}\cup \{ \infty \}.$
\end{center} 

Closely related to risk functionals are \gls{pcps} denoted by \gls{premium}, meaning the actual price the insurer charges for coverage of a risk. Often a PCP is explicitly derived from a risk measure. For example, indifference premiums are often derived by solving $H[\pi[X]-X]=0$, which (assuming translation invariance) simply yields $\pi = -H[-X]$. Aside from pricing, the calculation of risk capital is important for shareholder and management purposes such as solvency requirements (\cite{mcneil2005quantitative}). Infrequent but large losses in insurance and finance have led to the widespread adoption of tail-based measures of risk. Most prominent among these are \gls{VaR}, \gls{ES} and the closely related \gls{TCE}.

Let $X \in \chi$ have a \gls{cdf_words} given by $F_{X}(x)$. Given some prudence level $q\in (0,1)$,  the VaR is simply defined as the $q$-th quantile of the distribution of $X$:
$$ VaR_{q}[X]= x_q =\inf \left\lbrace x | F_{X}(x) > q \right\rbrace, $$

or simply $x_q =F_{X}^{-1}(q)$ for continous distributions. 

\noindent The ES attempts to capture the mean loss over a threshold by averaging the VaR over all prudence levels greater than or equal to $q$:
$$ ES_{q }[X]=\frac {1}{1-q}\int_{q}^{\infty }{VaR}_{\gamma }[X]d\gamma. $$

\noindent Finally, when $F_X$ is continuous the ES coincides with the more intuitive TCE, given by
$$TCE_{q}[X] = \expect[X \mid X > x_q ].$$

\noindent In other words, like ES, TCE measures the mean loss over some threshold for a given prudence level. The ES/TCE is often touted as an alternative to VaR because it is \textit{coherent} in the sense of Artzner  \cite{artzner1999coherent}. The VaR is however coherent for elliptical loss variables\footnote{This can easily be shown using properties of elliptical distributions and the triangle inequality.}. 

Given a risk measure $H$ and $n$ random variables $X^{(1)},...,X^{(n)}$  representing the total losses from $n$ individual business lines, the aggregate risk capital is $H[S]$ where $S= X^{(1)}+...+X^{(n)}$. The calculation of $S$ and $H[S]$ is the first step in any risk management framework and is mandatory under insurance and banking regulations (e.g.\ SolvencyII/Swiss Solvency Test, Basel III). 

Corresponding to each choice of risk measure are capital allocation rules. Once $H[S]$ has been computed, it is natural to ask how the individual $X^{(i)}$'s contribute to $H[S]$. Consider again a financial institution with $n$ business lines and corresponding aggregate loss $S=\sum_{i=1}^n X^{(i)}$. The capital allocated to each line is denoted $A[X^{(i)},S]$. Given $S$ and risk capital $H(S)$, the question is: how to appropriately calculate the capital allocations $A[X^{(i)},S]$ such that $H[S] =\sum_{i=1}^n A[X^{(i)} , S]$? For profitability testing and other internal analyses (e.g.\ cost sharing, pricing \cite{venter2004capital})  it is important to know which business lines contribute the most to aggregate risk.

For the TCE, there is an extremely natural choice of allocation rule. Taking advantage of the additivity of expectation, we have 

\begin{center}
$\expect[ S | S > s_q ] = \sum_{i=1}^n \expect[ X^{(i)} | S > s_q ].  $
\end{center}
Thus, given an aggregate loss that exceeds a prudence level, the $k^{th}$ allocation is its expected contribution to this excess. 

Generally, calculating allocations is very involved even when given a specific risk measure like the TCE. First there is the non-trivial task of determining the stochastic properties of $S$ (\cite{miles2019} and references therein). Second, in addition to the potentially complicated relationship between the $X^{(i)}$'s and $S$, there is often a dependence structure between the different $X^{(i)}$'s. That being said, in certain cases this problem can be reduced in complexity to determining the aggregate risk. A good example of this was first put forward by Panjer \cite{panjer2002measurement} for the TCE risk measure when bivariate Normal losses were assumed. In such a case allocations were shown to be linear in the aggregate risk capital:

\begin{center}
$A[X^{(i)},S]= \expect[ X^{(i)}] + \gamma_{X^{(i)}, S} \left(  H(S) - \expect[S]   \right)$
\end{center}


\noindent where $\gamma_{X^{(i)}, S}$ is the typical regression coefficient, namely $\cov[X^{(i)},S]/\var[S]$ where \gls{Covariance} and \gls{Variance} are the covariance and variance respectively. This is obviously reminiscent of exposure to systematic risks \textit{à la} CAPM in finance. The assumption of normality was heuristically justified by regarding business lines as the sum of enough individual policies to invoke the Central Limit Theorem (\gls{clt}). Later results showed this linear risk decomposition could be extended to more general elliptical models \cite{landsman2003tail}. More recently, \cite{furman2010general} and then \cite{furman2018weighted} have shown that such a decomposition is possible under very general considerations. This has led to the development of an insurance analogue to the CAPM known as the \gls{wipm} (\cite{furman2017beyond}). Crucially, \cite{furman2018weighted} shows that the linearity of elliptical conditional expectation is key in decomposing the allocations into linear functions of aggregate risk capital. There are however limitations to using elliptically distributed losses. For more general criticisms on the topic of elliptical linear dependence in the risk management literature, see  \cite{embrechts2002correlation} and \cite{bilodeau2004tail}. 


The instinct to use the \gls{clt} in risk aggregation is broadly correct. But in order to approximate $S$ via the classical CLT the following must be true: i) $n$ must be large, ii) the $X^{(i)}$'s must be sufficiently similar and iii) the moments of $X^{(i)}$'s must be ``well behaved" (i.e.\ $\var[ X^{(i)} ] < \infty$, small skewness, etc.). In practice, at least one of these usually fails. Insurance losses for example are often totally skewed and highly heterogeneous, and can have infinite second moments. In finance, there is a perennial thread in the literature disputing the use of Normal models (starting with \cite{mandelbrot1967variation}; for a more modern reference see \cite{rachev2011financial}). The inapplicability of Normal models may seem discouraging in applying the CLT but the Normal distribution is \textit{not} the only possible limit to schemes of sums of normalized random variables. The class of Stable distributions and the \gls{gclt} can address the abovementioned shortcoming. 

The organization of \ifbool{is_paper}{this paper}{Chapter \ref{ch: 3}} is as follows. In Sections \ref{sec: Stable} and \ref{SEC: Model} we will outline our proposed model along with the application of the GCLT. We will provide a more detailed background on the linear decomposition of allocations in Section \ref{SEC: WIPM} vis-à-vis Stable distributions. Section \ref{SEC:Stable TCE} will include a formula for the TCE in the Stable case and the corresponding allocations. We end the \ifbool{is_paper}{paper}{chapter} with some notes on how to compute the quantities related to stable distributions in Section \ref{SEC:Evaluation} followed by concluding remarks in Section \ref{SEC: Conclusion}.  All technical proofs are included in Appendix \ref{app: AppendixC}; for more background on Stable random variables see Appendix \ref{sec: ID}.

\section{Multivariate Stable Random Variables}\label{sec: Stable}

\ifbool{is_paper}{We}{Recall that we} can model the losses of the $n$ business lines of an insurance company by the vector $\mathbf{ X }=(X^{(1)},...,X^{(n)})$ and the total losses by $S= X^{(1)}+...+X^{(n)}$. The quantity $S$, being the sum of Stable marginals in a Stable vector, will also be Stable\footnote{The Stables share this property with the Normal class. However, in the same way that a vector with Normal marginals may not be distributed as a multivariate Normal (for a classic counterexample see \cite{melnick1982misspecifications}), Stable vectors have more properties than those of a vector with straightforward Stable marginals.}. Likewise, $(X^{(i)},S)$ will be a bivariate Stable vector with results that will ultimately make the calculations of risk capital and allocations relatively straightforward (Section \ref{sec: stable decomp}). Before moving on to the questions of risk capital and allocation, in this Section we describe some of the unique properties of Stable random vectors.  

Stable random vectors can be defined in a similar way to definition \ref{def: Stable_Def_1}. However, more involved results will require the introduction of some extra machinery. General Stable $d$-dimensional random vectors that preserve the essential properties of Stable distributions are determined by a standard shift vector $\boldsymbol{\mu}$ and a finite measure $\Lambda$ on the Borel sets of the $d$-dimensional unit sphere. The measure $\Lambda$ is often termed a ``spectral measure'' in the literature. This spectral measure determines the dependence structure of the random vector's components. As in the univariate case, in only a few cases will a closed-form pdf of the distribution actually exist. 

\begin{definition}[Stable Random Vector]\label{Stable_multi_def} $\mathbf{ X } \in \mathbb{R}^d$ is a Stable vector if it  has characteristic function
\begin{equation}\label{def: multi_CF}
\phi_{ \mathbf{ X }|\mu, \Lambda }\left( \bm{\tau} \, \right) 
=\expect\left[ \exp\{  i \inprod{\mathbf{ X }}{ \bm{\tau} } \} \right]
= \exp \left\lbrace -\int_{S_{d}} \Upsilon_\alpha( \inprod{\bm{\tau}}{ \mathbf{ s } } ) \Lambda(d \mathbf{ s })  + i\inprod{  \mathbf{ \mu } }{ \bm{\tau}}  \right\rbrace ,
\end{equation}
denoted $\mathbf{ X }\backsim S_{\alpha}(\Lambda ,\mathbf{ \mu })$, where 

\begin{enumerate}
\item $\Lambda$ is a spectral measure on the unit sphere $S_{d}$ determining the dependence structure and distribution of $\mathbf{ X }$

\item $\boldsymbol{\mu}$ is a standard location vector

\item $\Upsilon_\alpha$ is the CF of a totally skewed univariate stable variable (see \ref{eq: Stable_CF}):

\[  \Upsilon_\alpha(u) =  \left\{
\begin{array}{ll}
      |u|^\alpha(1- i a \,\sign(u)) & \alpha \neq 1 \\
      |u|(1 + i \frac{2}{\pi} \,\sign(u) \ln(u))  & \alpha = 1\\
\end{array} 
\right. \]

\end{enumerate}
\end{definition}

Stable random vectors defined by (\ref{def: multi_CF}) do share a nice property with Normals and Normal-scale mixtures: linear combinations of their marginals are univariate Stable. Let $\boldsymbol{\mu}=0$ and define the following:
\begin{align}
\sigma^\alpha\left(\bm{\tau}\,\right) &= \int_{S_{d}} |\inprod{\bm{\tau}}{\mathbf{ s }}|^\alpha \Lambda(d \mathbf{ s }) \label{eq: sigma_def} \\
\beta\left(\bm{\tau}\,\right) &= \frac{1}{ \sigma^{\alpha}\left(\bm{\tau}\,\right) } \int_{S_{d}} \sign(\inprod{\bm{\tau}}{ \mathbf{ s } })|\inprod{\bm{\tau}}{ \mathbf{ s } } |^\alpha \Lambda(d \mathbf{ s }) \label{eq: beta_def}\\
I_{\mathbf{ X }}\left(\bm{\tau}\,\right)  &=  \left\{
\begin{array}{ll}
      \sigma^\alpha\left(\bm{\tau}\,\right)(1-i\beta\left(\bm{\tau}\,\right)\tan(\frac{\pi \alpha}{2})) & \alpha \neq 1 \\
      \sigma\left(\bm{\tau}\,\right)\left( 1-i\int_{S_{d}} \inprod{\bm{\tau}}{\mathbf{ s }} \ln(\inprod{\bm{\tau}}{\mathbf{ s }}) \Lambda(d \mathbf{ s }) \right)  & \alpha = 1.\\
\end{array} 
\right.  
\end{align}
Then $\inprod{\bm{\tau}}{\mathbf{ X }}$ is a one-dimensional random variable with characteristic function
$$\mathbf{E} \exp\{  i u\inprod{ \mathbf{ X }}{ \bm{\tau} } \} = \exp \lbrace -I_{\mathbf{ X }}\left(u \bm{\tau}\,\right) \rbrace. $$ 
 
\vspace{3 mm} 
 
Breaking from the multivariate Normal, the converse is not always true:

\begin{theorem}[\cite{samoradnitsky_book}]\label{Samoradnitsky}

Let $\mathbf{ X }$ be a random vector in $\mathbb{R}^N$. If $\forall \mathbf{t} , \inprod{ \mathbf{t} }{ \mathbf{ X }}$ is stable with $\alpha>1$ then $\mathbf{ X }$ is a Stable vector.

\end{theorem}

Obviously, a potential issue for the \textit{practical} usage of Stable random vectors is the specification of the dependence structure and therefore the spectral measure $\Lambda(\cdot)$. Fortunately, the spectral measure can be naturally approximated by a much simpler discrete object to arbitrary precision: 
\begin{definition}[Discrete Spectral Measure]

Given a set of points $\mathbf{ s }_i \in S_{d}$ and a corresponding set of weights $\gamma_i>0$,  we can define a spectral measure $\Lambda : S_d \rightarrow \mathbb{R}^{+}$ as follows:
\begin{equation}\label{SM_D}
\Lambda(\cdot) = \sum^{m}_{i=1} \gamma_i \delta_{\mathbf{ s }_i}(\cdot).
\end{equation}

\end{definition}
For a proof and means of construction of this measure, see \cite{byczkowski1993approximation}. Given that discrete measures appear naturally in many contexts (\cite{nolan2003book}) and are much easier to manipulate, we will simply consider them on their own. Additionally, discrete measures give rise to a powerful stochastic representation. A nice feature of elliptical distributions is the ease with which they can be manipulated under linear transformations. Fortunately, this is still true when a Stable vector is described by a discrete spectral measure.

Assume $\mu=0$ and define $Z^{(j)}\sim S_\alpha(1,1,0)$ i.i.d.\ and 

\begin{equation}\label{Stoch_Rep}
  \mathbf{ X } := \left\{
\begin{array}{ll}
       \sum_{j=1}^{n}  \mathbf{ s }_j \gamma_j^{1/ \alpha} Z^{(j)} & \alpha \neq 1 \\
       \sum_{j=1}^{n} \mathbf{ s }_j  \gamma_j (Z^{(j)} + \frac{2}{\pi} \ln \gamma_j ) & \alpha = 1.\\
\end{array} 
\right. 
\end{equation}

It is easy to show that the characteristic function of $\mathbf{ X }$ is
\begin{equation}\label{CF_D}
\phi_{ \mathbf{ X }}( \bm{\tau} ) = \exp \left( - \sum_{i=1}^n  \gamma_i \Upsilon_\alpha(\inprod{ \bm{\tau} }{ \mathbf{ s_i } }) \right).
\end{equation}
That is, $X$ has a characteristic function of the form (\ref{CF_D}) if the L\'{e}vy measure in (\ref{def: multi_CF}) is a discrete measure. Once again we will exclusively look at the $\alpha \ne 1$ case. Without loss of generality we will include in the set of $(\gamma, \mathbf{ s })$ the twin pairs $(\gamma_i,\mathbf{ s_i })$ and $(\gamma_{-i},\mathbf{s_{-i}})=(\gamma_{-i},-\mathbf{s_{i}})$, so that 
\begin{align*}
\mathbf{ X } &= \sum_{j=1}^{2n} \gamma_j^{1/ \alpha}  \mathbf{ s }_j Z^{(j)} \\
&=[\mathbf{ s }_1 \dots \mathbf{ s }_{n}, -\mathbf{ s }_1 \dots -\mathbf{ s }_n ]\text{diag}[ \gamma_{+1},\dots,\gamma_{+n},\gamma_{-1} \dots, \gamma_{-n}]^\frac{1}{\alpha} \mathbf{Z} \\
&=\mathbf{S}_{d \times 2n}\mathbf{D}_{2n \times 2n}^\frac{1}{\alpha} \mathbf{Z}.
\end{align*}

If $\gamma_{i}=\gamma_{-i}$ then we will say that $\Lambda$ is \textit{symmetric}. It is tempting to look at the expression $\mathbf{ X }=\mathbf{S}_{d \times 2n}\mathbf{D}_{2n \times 2n}^\frac{1}{\alpha} \mathbf{Z}$ and conclude that  $\mathbf{ X }$ is elliptical if $\Lambda$ is symmetric. In general, however, this will not be true: 

\vspace{- 5mm}

\begin{align*}
\mathbf{ X } &= \sum_{j=1}^{2n} \mathbf{ s }_j( \gamma_{+j}^{1/ \alpha} Z^{+j} - \gamma_{-j}^{1/ \alpha} Z^{-j}) \\ 
&= \dfrac{1}{2^{1/\alpha}}\sum_{i=1}^{n} \mathbf{ s }_i Z_i^{sym} \\
&= \dfrac{1}{2^{1/\alpha}}\ [\mathbf{ s }_1 \dots \mathbf{ s }_{n}] \mathbf{Z}^{sym}.
\end{align*}

Each marginal in $\mathbf{Z}^{sym}$ has $\beta=0$. Only if $[\mathbf{ s }_1 \dots \mathbf{ s }_{n}]$ is a Cholesky factorization of a positive definite matrix do we have an elliptical Stable vector. 

We conclude this section on Stable vectors by stating a theorem that will be highly relevant in the following sections. 

\begin{theorem}\label{regression} (\cite{samoradnitsky_book}) Let $(X_2,X_1)\backsim S_{\alpha}(\Lambda ,0 )$ be a jointly distributed Stable random vector. Then 

$$ \Cexpect{X_2}{ X_1 =x } = \kappa_{2,1}x + a \sigma_1^\alpha( \lambda_{2,1} - \beta_1 \kappa_{2,1}) \frac{h(x)}{ \pi f_{X_1}(x)}$$
where 
\vspace{-3 mm}
\begin{align*}
\kappa_{2,1} &= \frac{[X_2,X_1]}{\sigma_1^\alpha} = \frac{\int_{S_{d}}  s_2 |s_1|^{\alpha-1} \text{\normalfont sign}(s_1) \Lambda(d \mathbf{ s })}{\sigma_1^\alpha} \\
\lambda_{2,1} &= \frac{ \int_{S_{d}}  s_2 |s_1|^{\alpha-1} \Lambda(d \mathbf{ s })}{\sigma_1^\alpha} \\
h(x) &= \stablefun{t^{\alpha-1}}{x}{1}
\end{align*}
and $f_{X_1}(x)$ is the pdf of $X_1$. 

\end{theorem}

\section{Multivariate Stable Insurance Losses}\label{SEC: Model}

\ifbool{is_paper}{As mentioned in \ref{sec: p1 intro}, the symmetric nature of the normal makes it awkward for insurance applications. In this paper we generalize the multivariate normal to the totally skewed Stable family of distributions. Considering the data complied by \cite{eaton1971extreme} and \cite{embrechts2013modelling} we believe this makes for a more realistic application to insurance.}{In Section \ref{subsec: stable abrm} we discussed briefly how and why a multivariate Stable model may be a natural choice for heavy-tailed insurance data.} Given that it is a central object of study in this work we will take some time to examine this model and its possible drawbacks more fully. Ultimately we want to be able to approximate $N$ business lines with a multivariate Stable vector. Consider an insurance company and assume for simplicity that it has two business lines. Further suppose each line sells identical policies whose losses can be (at least asymptotically) described as Pareto-tailed random variables $L^{(1)}$ and $L^{(2)}$ with tail parameters $\alpha_1$ and $\alpha_2$ respectively\footnote{ Specifically, $1-F_X(x)\sim \frac{1}{x^\alpha}$ as $x\rightarrow\infty$ and $F_X(x) \sim \frac{1}{|x|^\alpha}$ as  $x \rightarrow -\infty$.}.

Assuming a roughly equal number of policies, $n$, are sold for each line, we want to model the total aggregate loss random vector: 
$$\mathbf{ X }_n= \sum_{i=1}^{n} {\begin{pmatrix} L_{i}^{(1)}\\\ L_{i}^{(2)}\\\end{pmatrix}}$$ 
 
Obviously we would like to show that we can model $\mathbf{ X }_n$ as a Stable random \textit{vector}. As we will see, however, the only case in which we can do so without introducing degenerate marginals is when $\alpha_1 = \alpha_2$. Since $\alpha>1$ we can set $q_n =  (n^{1-\frac{1}{n^\alpha}} )(\mu_1,\mu_2)^T$ where $\mu_i$ are the respective means. Consider the linear combinations of the centred normalized losses:
\begin{align*}
\frac{1}{n^{1/\alpha}} \inprod{ \bm{\tau} }{ \mathbf{ X_n } } -q_n &=  \frac{\sum_{i=1}^n t_1 (L_{i}^{(1)}-\mu_1)  + t_2 (L_{i}^{(2)} - \mu_1) }{n^{1/\alpha}} \\
 & =  \frac{\sum_{i=1}^n (t_1 L_{i}^{(1)} + t_2 L_{i}^{(2)} ) - ( t_1\mu_1 + t_2 \mu_2) }{n^{1/\alpha}} .
\end{align*}

To make sense of the above expression we will make use of a result of \cite{tucker1968convolutions}. 

\begin{theorem}[\cite{tucker1968convolutions}, Lemma 3]\label{thm: Tucker}
Consider two random variables $Z_1$ and $Z_2$ with Pareto tails of index $\alpha_1$ and $\alpha_2$. We have that
$$  \frac{1 - F_{Z_1+Z_2}(k x)}{1 - F_{Z_1+Z_2}(x)} \sim x^{- \min{ \{ \alpha_1,\alpha_2}\} }\text{  as  } k \rightarrow \infty.$$
\end{theorem}

This theorem implies that $t_1 L_{i}^{(1)} + t_2 L_{i}^{(2)}$ will have tail index $\alpha$, and as $n \rightarrow \infty$, by the GCLT we have
\begin{equation}\label{ell}
\frac{1}{n^{1/\alpha}}\inprod{ \bm{\tau} }{  \mathbf{ X }_n } -q_n  \sim S_\alpha( \sigma(\tau), \beta(\tau), \sigma(\tau) ).
\end{equation}

\noindent This is true for any linear combination and so $\frac{1}{n^{1/\alpha}} \mathbf{ X }_n -\mathbf{ q }_n$ weakly converges to a Stable random vector by Theorem \ref{Samoradnitsky} and the Cramer--Wold theorem. 

Assume now that $\alpha_1 \neq \alpha_2$. In this case the tail index of $\inprod{ \bm{\tau}}{ \mathbf{ X_n }}$ will depend on $\bm{\tau}$. For $\bm{\tau}$ where $t_2 , t_1 \neq 0$ the previous picture does not change significantly; by Theorem \ref{thm: Tucker} the tail of $t_1 \ell_{i}^{(1)} + t_2 \ell_{i}^{(2)}$ will be $\alpha^* = \min\{ \alpha_1, \alpha_2 \}$ and we will still have the situation in (\ref{ell}). 

However, say $\bm{\tau} = (0,1)^T$ then
\begin{align}\label{degen1}
\frac{1}{ n^{1/{\alpha^*}} }\inprod{ \bm{\tau} }{\mathbf{ X_n } }  - q_n  
& = \frac{\sum_{i=1}^n  L_{i}^{(2)}  -  \mu_2 }{n^{1/{\alpha^*}}} \\
&= \frac{1}{n^{ \frac{1}{\alpha^*} - \frac{1}{\alpha_2} }}  \frac{\sum_{i=1}^n  \ell_{i}^{(2)}  -  \mu_2 }{n^{1/{\alpha_2}}}\\
& \sim S_{\alpha_2}( n^{ \frac{1}{\alpha_2} - \frac{1}{\alpha^*} } \sigma , \beta , 0) \\
& \rightarrow \delta(x)\label{degen2}
\end{align}
and indeed, the proof of Theorem \ref{Samoradnitsky} in \cite{samoradnitsky_book} begins by showing that if all linear combinations of a random vector are Stably distributed then those that are non-degenerate have the same tail index! 

In light of this we could envision (for $N$ business lines) a model of the following kind:
\begin{equation}\label{eq: expanded loss model}
\mathbf{ X }_n = \sum_{j=1}^n \begin{pmatrix} a^{(1)} I_j^{(1)} + b^{(1)}M_j\\ \vdots\\ a^{(N)} I_j^{(N)} + b^{(N)}M_j \end{pmatrix}
\end{equation}
where we have not yet specified our ``idiosyncratic'' and ``market'' factors, the $I_j^{(i)}$'s and $M_j$'s, beyond the fact they are Pareto-tailed. As previously mentioned, according to \cite{embrechts2013modelling} and others many insurance losses exhibit Pareto-tailed behaviour over a certain threshold. We believe it is prudent from a risk management perspective to choose the tail index of losses to be the minimum such index observed across all business lines. This also allows us to approximate the losses the insurance company faces as a Stable vector \textit{without} the degeneracies encountered in (\ref{degen1})--(\ref{degen2}).

Take all the tail indexes of the losses to be $\alpha$. Abusing notation, the GCLT implies that
$$\frac{1}{n^{1/\alpha}}\inprod{ \bm{\tau}}{  \mathbf{ X_n } } \longrightarrow  \inprod{\bm{\tau}}{ \mathbf{ X }} \sim S_{\alpha}(\sigma\left(\bm{\tau}\,\right),\beta\left(\bm{\tau}\,\right),\mu\left(\bm{\tau}\,\right)).  $$

If this is true for all $\bm{\tau}$ and $\alpha>1$ then $\mathbf{ X }$ is a Stable vector by Theorem \ref{Samoradnitsky}, and by the Cramer-Wold theorem we have
$$\frac{1}{n^{1/\alpha}}\mathbf{ X }_n \longrightarrow \mathbf{ X } = \bm{a} \circ \mathbf{Y} + \bm{b} Z $$



\noindent where we can recover $\mathbf{Y}$ and $Z$ from the CGLT convergence of the $I_j$ and $M_j$ sums. \ifbool{is_paper}{}{This is the model we introduced in Section \ref{eq: ABRM vector}. } 


\section{Weighted Insurance Pricing}\label{SEC: WIPM}

\subsection{Weighted Risk Measures }

Consider a loss random variable $X$. We can compute its expectation using the inverse CDF and integrating over probabilities:

\begin{equation*}
\expect[ X] = \int_{0}^1 F^{-1} (p) dp
\end{equation*}

In order to avoid ruin with probability one, insurers require net premiums to be at least $\expect[X]$. The easiest way to do this is to distort probabilities of events in such a way as to guarantee that the net premiums satisfy this lower bound. That is, we calculate the expectation or net premium under a distorted distribution. This is commonly achieved through a \textit{distortion function} $ g : [0,1] \rightarrow [0,1]$, an increasing function such that $g(0)=0$ and $g(1)=1$. Define for the net premium the class of risk measures 
\begin{equation*}
H[X,g] = \int_{0}^1 F^{-1} (p)  g'(1-p) dp
\end{equation*}
called distortion risk measures (see \cite{balbas2009properties}). Note that $g'(1-p)$ is non-negative and non-increasing: large losses are emphasized and lossless scenarios are de-emphasized. These distortion risk measures encompass a large class of well-studied risk functionals and corresponding premium calculations.  

A similar procedure for achieving the same goal of reweighing the loss probabilities is to directly re-weight the distribution function. Given a random variable $S$ and a weight function $w$ such that $0 < \expect[w(S)] < \infty$, we can define the CDF of the weighted distribution as
\begin{center}
$ F_{w;S}(s) = \dfrac{\expect[ \mathbf{1} \lbrace S \leq s \rbrace  w(S)]}{\expect[w(S)]}$. 
\end{center}

We can define the weighted risk measures similarly to our definition of the distortion class, as an expectation with respect to the new distribution: 
\begin{equation*}
H_w[S] = \dfrac{\expect[ S w(S) ]}{\expect[ w(S) ]}.
\end{equation*}  

Note that the class of distortion risk measures are a special case\footnote{ $ w(s) = g'( \bar{ F }_S (s) )$}. Additionally, while we will primarily be interested in the case of the TCE ($w(s) = \mathbf{1} \lbrace s > s_q \rbrace$), this class easily recovers other standard risk measures (see \cite{FURMAN2008263,FURMAN2008459}).

\subsection{Weighted Allocations}\label{A_derivation}

Perhaps the most important and useful property of the class of weighted risk measures is the ease with which corresponding allocation rules can be derived and interpreted. To simplify the analysis, we assume continuous risks, so that
\begin{equation*}
dF_{w;S}(s)= \frac{w(s)}{\expect[w(S)]} f_S(s) dx .
\end{equation*}

Recall the notation that assigns, for a financial institution with $n$ business lines, losses $X^{(1)}, X^{(2)},...,X^{(n)}$ and aggregate loss $S=\sum_{i=1}^N X^{(i)}$. Using additivity of expectation, 
\begin{align*}
\dfrac{\expect[ S w(S) ]}{\expect[ w(S) ]} 
&=\sum_{i=1}^n \dfrac{\expect[ X^{(i)} w(S) ]}{\expect[ w(S) ]} \\
&=\sum_{i=1}^n \int_{} \int X^{(i)} \dfrac{ w(S) }{\expect[ w(S) ]} f_{(X,S)}(x,s) dxds \\
&=\sum_{i=1}^n \int \left[ \int X^{(i)} f_{X|S}(x|s) dx \right] \dfrac{ w(S) }{\expect[ w(S) ]} f_{S}(s) ds \\
&=\sum_{i=1}^n \int \expect[ X^{(i)} |\, S =s ]\, dF_{w;S} (s) .\\
\end{align*}

One can show that the quantity $\int \expect[ X^{(i)} |\, S =s ]\, dF_{w;S} (s) $ satisfies many properties desired in an allocation rule: no unjustified loading, consistency and of course full additivity. Furthermore \cite{FURMAN2008263} shows that it is non-negative and no undercut holds in the TCE case. To that end, we define for a given weight function $w$ the allocation 
$$ A_w[X,S] = \frac{ \expect[X^{(i)} w(S)] }{ \expect[ w(S) ] }=\int \expect[ X^{(i)} |\, S =s ]\,  dF_{w;S} (s). $$
This allocation is easily interpretable in the case that $f_{S},f_{X} \in L^2$. Assume an insurer's preferences or utility for profit and losses is quadratic for each business line. This is a standard assumption in many basic versions of various financial models such as the CAPM \cite{panjer1998financial}. We have
\begin{align*}
\min_{a_i} \expect\left[ (X^{(i)}-a_i)^2 \frac{w(S)}{\expect[w(S)]} \right] 
& = \min_{a_i} \expect \left[ \expect \left[ (X^{(i)}-a_i)^2  \Big\vert S \right] \frac{w(S)}{\expect[w(S)]}  \right] \\
&= \min_{a_i} \int \expect \left[ (X^i-a_i)^2 \Big\vert S=s \right] dF_w(s).
\end{align*}

We can easily prove that:
\begin{align*}
a_i &= \int \expect[ X^{(i)} |\, S =s ]\, dF_w (s) .\\
\end{align*}

\subsection{Weighted Allocations given Stable Losses}\label{sec: stable decomp}

Having elucidated the necessary properties of both weighted allocations and Stable vectors, we are now ready to derive results for weighted allocations in the Stable case. We begin by referring back to Section \ref{A_derivation} and specifically the derivation of $A_w[X^{(i)}, S]$. Making use of Theorem \ref{regression} we can find a very general form for allocations when the joint losses $X^{(i)}$ are described by a Stable random vector:  
\begin{align*}
\expect[w(s)] A_{w}[X^{(i)}, S ] &= \int \Cexpect{X^{(i)}}{ S=s } w(s) f_{S}(s) ds \\
&= \int \left(\kappa_i s + a \sigma_S^\alpha( \lambda_i- \beta_S \kappa_i ) \frac{h(s)}{f_S(s)}\right) w(s) f_{S}(s) ds \\
&= \kappa_i  \int s  w(s) f_{S}(s) ds + a \sigma_S^\alpha(  \lambda_i- \beta_S \kappa_i ) \int \frac{h(s)}{f_S(s)} w(s) f_{S}(s) ds .
\end{align*}

Where $\kappa_i=\frac{[X_i,S]}{\sigma_S^\alpha}$ and $\lambda_i$ is similarly defined with respect to $X_i$ and $S$. This gives us that for \textit{any} appropriate weight function,
\begin{equation}\label{Stable_WIPM}
  \boxed{A_{w}[X^{(i)}, S ] = \kappa_i  H_w(S) + a \sigma_S^\alpha( \lambda_i- \beta_S \kappa_i ) \frac{ \int h(s) w(s)  ds}{\expect[w(S)]}. }
\end{equation}

Before continuing, it is worth making a couple of observations about Eq.\ (\ref{Stable_WIPM}). While we can simplify the task of calculating allocations to the evaluation of the aggregate risk capital $H_w(S)$ and some skewness term (the integral involving $h(s)$), this in itself is not a trivial task.  Surprisingly, as we will see, the skewness term disappears, leaving just the risk capital term. That being said, there are in general not many well-known results for risk measures involving Stable Losses. Only \textit{numerical} results are known on $H_w(S)$ for Stable $S$ in the TCE case \cite{stoyanov2006computing}. Specifying Eq.\ (\ref{Stable_WIPM}) for the TCE ($w(x)=\mathbf{1}_{(x>x_q)}$) case will be our focus for the rest of this paper. 

\paragraph{As Applied to the Model}

The quantity $a \sigma_S^\alpha\frac{ \int h(s) w(s)  ds}{\expect[w(s)]}$ is shared for all $X^{(i)}$:  the effects of skewness enter the allocation only in the $\lambda_i- \beta_S \kappa_i$ term. Interestingly, our model allows for a non-elliptical dependence while preserving the results of \cite{landsman2003tail}. If the losses $L_i^{(j)}$ are Pareto-tailed and totally skewed then $1-F_{S}(s) \sim C s^{-\alpha}$ and $F_{S}(s) \sim 0$. In the limit when approximated by a Stable, the marginals will be totally skewed to the right (see Theorem \ref{thm: GCLT}). Assume a discrete spectral measure for the vector $(X_1,...,X_N,S)^{T}$ where for $\mathbf{ s }_j \in S_{N+1}$ we have $\mathbf{ s }_j = (s_j^{(1)},...,s_j^{(i)},...s_j)$. Recalling (\ref{eq: sigma_def}) and (\ref{eq: beta_def}), this yields
\begin{align*}
\sigma_S^\alpha &= \sum_{j=1}^m |s_j|^\alpha \gamma_j \\
\beta_S &= \frac{1}{\sigma_S^\alpha} \sum_{j=1}^m |s_j|^\alpha \text{sign}(s_s^{(j)}) \gamma_j.\\
\end{align*}

Clearly, there is no nonzero support in the spectral measure where $\text{sign}(s_s^{(j)})= -1$ and
$$ \kappa_i = \frac{1}{\sigma_S^\alpha}\sum_{j=1}^m  s_j^{(i)}|s_j|^{\alpha-1} \text{sign}(s_s^{(j)})\gamma_j = \frac{1}{\sigma_S^\alpha}\sum_{j=1}^m  s_j^{(i)}|s_j|^{\alpha-1}\gamma_j = \lambda_i.$$
So then $(\lambda_i- \beta_S \kappa_i)=0$, and in the model,
\begin{equation}\label{Model_WIPM}
A_{w}[X^{(i)}, S ] = \kappa_i  H_w[S].
\end{equation}

Recall that Theorem \ref{regression} considered the case with no location parameters (or means where they exist). One can easily add them in: 
\begin{equation}\label{Model_WIPM2}
A_{w}[X^{(i)}, S ] = \expect[X^{(i)}] + \kappa_i(  H_w[S] - \expect[S]).
\end{equation}

Calculating the allocations in this case will only require us to calculate $\kappa_i$ and $H_w(S)$. In the next section we do just that, providing a result for the TCE in the Stable case and a simple example involving $\kappa$. 

\section{Stable TCE and Portfolio Risk Allocation}\label{SEC:Stable TCE}

In this section we will provide a representation of the TCE in the Stable case using the Fox H-functions described in Appendix \ref{app: AppendixB} to represent the Stable pdf. We will use two lemmas proved in Appendix \ref{app: AppendixC}. It is worth noting that in \cite{stoyanov2006computing} a formula for the Stable TCE is developed through direct numerical integration, whereas the Fox H-function representation used in this work allows us to leverage the numerical convenience therein (see Section \ref{SEC:Evaluation}).


Following the derivation of this representation, we will work with our simple example of an insurance company, using \ref{thm: TCE} and (\ref{Model_WIPM}) to compute the allocations. 
Given that $H_w[S]$ is the TCE when  $w(s)= \mathbf{1}_{s>s_q}$, then 
$$  H_w[S]=  \frac{1}{1-q} \int_{s_q}^\infty s f_{S}(s) ds$$
where $\expect[w(s)] = \int_{s_q}^\infty  f_{S}(s) ds = 1- u $ and, given the cdf $F_S (s)= P(S \leq s)$, we have $s_q = F_S^{-1}(q)$. To compute such an integral we will need to represent $f_{S}(s)$ as the inverse \glssymbol{fourier}[Fourier transform] of the characteristic function (\ref{eq: Stable_CF}), entailing a double integral. In order to do this, will need to state a few results, beginning with expressing the following \glssymbol{laplace}[Laplace transform] in terms of a simple H-function.

\begin{lemma}[The Laplace transform of $t^j e^{-bt^\alpha}$]\label{Lemma 1} 
$$\lap{x}{t^j e^{-bt^\alpha} }= \frac{1}{\alpha b^\frac{j+1}{\alpha}} H_{{1,1}}^{{\,1,1}}\!\left[ \frac{x}{ b^\frac{1}{\alpha}} \left|{\begin{matrix} \left( 1-\frac{j+1}{\alpha},\frac{1}{\alpha}\right) \\(0,1)\end{matrix}}\right.\right]. $$
\end{lemma}

When applying Lemma \ref{Lemma 1} we shall often need to take the real part of an H-function with complex arguments. To that end we shall need the following:


\begin{lemma} \label{Lemma 2} 

For $z_1, z_2 \in \mathbb{C}$, $\nu_1 \in \mathbb{R}$ and $\nu_2 \in \mathbb{R}^+$,
\begin{align*}
&\hspace{5 mm} z_1^{\nu_1}
H_{{1,1}}^{{\,1,1}}\!\left[ z_2^{\nu_2} x  \left|{\begin{matrix} \left( a_1,A_1\right) \\(b_1,B_1)\end{matrix}}\right.\right] \\
&=\pi r_1^{\nu_1}  \left( H_{{2,2}}^{{\,1,1}}\!\left[ r_2^{\nu_2} x  \left|{\begin{matrix} \left( a_1,A_1\right) & \left( \frac{1}{2}-\frac{\theta_1 \nu_1}{\pi},{\frac{\theta_2 \nu_2}{\pi}}\right)  \\ (b_1,B_1) &  \left( \frac{1}{2}-\frac{\theta_1 \nu_1}{\pi},{\frac{\theta_2 \nu_2}{\pi}}\right)  \end{matrix}}\right.\right] + i H_{{2,2}}^{{\,1,1}}\!\left[ r_2^{\nu_2} x  \left|{\begin{matrix} \left( a_1,A_1\right) & \left(1- \frac{\theta_1 \nu_1}{\pi},{\frac{\theta_2 \nu_2}{\pi}}\right)  \\ (b_1,B_1) &  \left( 1-\frac{\theta_1 \nu_1}{\pi},{\frac{\theta_2 \nu_2}{\pi}}\right)  \end{matrix}}\right.\right]  
\right)
\end{align*}  
where $r_i = |z_i|$ and $\theta_i = \arg(z_i)$.  
  
\end{lemma}

\begin{theorem}\label{thm: TCE}

Let $S \sim S_\alpha(\sigma,\beta,0) $,  $s\geq 0$ , $r=\sqrt{1+(a \beta)^2}$, $\phi =\tan^{-1}(a \beta)  $ and $\gamma = \frac{1}{2} - \frac{\phi}{ \alpha \pi }$. Then the TCE is given by

$$TCE_{q}[S] = \frac{\sigma r^{\frac{1}{\alpha}} H_{{2,2}}^{{\,1,1}}\!\left[\frac{s_q}{\sigma r^{\frac{1}{\alpha}}}  
\left|{\begin{matrix}
(1-\frac{\alpha-1}{\alpha} , \frac{1}{\alpha})&(\gamma,\gamma)\\
(0,1)&(\gamma,\gamma)
\end{matrix}}\right.\right]}{1- q} $$ 

\end{theorem}

\begin{proof}

We need to evaluate the aforementioned double integral:

\begin{align*}
\int_{s_q}^\infty s f_{S}(s) ds &= \int_{s_q}^\infty s \left[ \frac{1}{2\pi} \int_{-\infty}^\infty e^{-its} \phi_{X} (t) dt \right] ds \\ 
&= \frac{1}{\pi} \int_{s_q}^\infty s \operatorname {Re}\left[ \int_{0}^\infty e^{-its} e^{-t^\alpha \xi} dt \right] ds \\ 
&= \frac{1}{\pi} \int_{s_q}^\infty s \operatorname {Re} \left[ \lap{is}{ e^{-t^\alpha \xi} } \right] ds \\
&= \operatorname {Re} \left[ \frac{1}{\pi} \int_{s_q}^\infty s \lap{is}{ e^{-t^\alpha \xi} } ds \right] \\
&\hspace{2.0 cm} \text{ (given the exterior integral is real) }   \\
&= \operatorname {Re} \left[ \frac{\alpha \xi}{\pi}  \left. \lap{is}{ t^{\alpha-2} e^{-t^\alpha \xi} }   \right\rvert_{s=\infty}^{s=s_q}  \right]. \\
\end{align*}

Recall Lemma \ref{Lemma 1}:
$$\lap{(is)}{ t^{\alpha-2} e^{-t^\alpha \xi} } 
= \frac{1}{\alpha \xi^{\frac{\alpha-1}{\alpha}}} H_{{1,1}}^{{\,1,1}}\!\left[ \frac{is}{\xi^\frac{1}{\alpha}}  \left|{\begin{matrix} \left( 1-\frac{\alpha-1}{\alpha},\frac{1}{\alpha}\right) \\(0,1)\end{matrix}}\right.\right]. $$

Fortunately, $H_{{1,1}}^{{\,1,1}}\xrightarrow[  ]{z\rightarrow \infty}0$\footnote{Refer to the definition of the Fox H-function in Appendix \ref{SEC:H_ref}. As the Bromwich path will have all positive real parts, then we can show that the integrand goes to zero and use the Dominated Convergence Theorem.  }, so 
$$\operatorname {Re} \left[ \frac{\alpha \xi}{\pi}  \left. \lap{is}{ t^{\alpha-2} e^{-t^\alpha \xi} }   \right\rvert_{s=\infty}^{s=s_q}  \right] = \sigma r^{\frac{1}{\alpha}} H_{{2,2}}^{{\,1,1}}\!\left[\frac{s_q}{\sigma r^{\frac{1}{\alpha}}}  
\left|{\begin{matrix}
(1-\frac{\alpha-1}{\alpha} , \frac{1}{\alpha})&(\gamma,\gamma)\\
(0,1)&(\gamma,\gamma)
\end{matrix}}\right.\right].$$

\end{proof}

While this characterization may seem uninformative, the Fox H-functions are a class of special functions with many known results. Furthermore, we will present a few natural ways of computing H-functions in Section \ref{SEC:Evaluation}.

\section{Evaluating Fox H-Functions}\label{SEC:Evaluation}

At this point it is natural to ask how to make use of the H-function representations in Section \ref{SEC:Stable TCE} and Appendix \ref{sec:pdf_proof}. In this section we will detail how one can compute the density function of the univariate Stable distribution via its H-function representation. The same methods can be repurposed for the evaluation of the TCE and other quantities found in Section \ref{SEC:Stable TCE} without any major changes.

There are two approaches to consider:

\begin{enumerate}

\item Numerically invert the integral transform defining the H-functions.

\item Find an equivalent series representation. 
\vspace{-5mm}
\end{enumerate}
\noindent \paragraph{Integral Transform Inversion} We will start with the more straightforward approach. Consider the Stable pdf as an H-function:
$$ f_X(x)= \frac{1}{\alpha \pi \sigma r^{\frac{1}{\alpha}} }H_{{2,2}}^{{\,1,1}}\!\left[\frac{x}{\sigma r^{\frac{1}{\alpha}} }  
\left|{\begin{matrix}
(1-\frac{1}{\alpha} , \frac{1}{\alpha})&(1 -  \gamma,\gamma)\\
(0,1)&(1-  \gamma,\gamma)
\end{matrix}}\right.\right]. $$ 

Interpreting this expression as an inverse \glssymbol{mellin}[Mellin transform] and simplifying the integrand using the Euler reflection formula gives 
\begin{equation}\label{integral}
f_X(x)  = \frac{1}{2 \pi i} \frac{1}{\alpha \pi \sigma r^{\frac{1}{\alpha}}} \int_{c-i \infty}^{c+i\infty} \G{s} \G{ \frac{1-s}{\alpha} } \sin( \pi [ \gamma - \gamma s ] ) (\frac{x}{\sigma r^{\frac{1}{\alpha}}})^{-s} ds .
\end{equation}

In order for the integral in (\ref{integral}) to converge, the path of integration must separate the poles of the two Gamma functions in the integrand (it being a Mellin--Barnes integral). The poles of the Gamma functions in (\ref{integral}) are $ s= -k_1$ and $s=1+\alpha k_2$ for $k_1,k_2 \in \mathbb{Z}^+$. The path of integration will be the line in the complex plane running from $ c- i\infty$ to $c+ i\infty$ with $c \in (0,1)$. In this case, (\ref{integral}) corresponds to the usual definition of an inverse Mellin transform.

The transformation $s' = \frac{s-c}{i}$ yields 
\begin{equation} \label{integral2}
f_X(x) = \frac{1}{2 \pi } \frac{1}{\alpha \pi} \int_{- \infty}^{\infty} \G{c+ is' } \G{ \frac{1- c - i s' }{\alpha} } \sin( \pi [ \gamma - \gamma (c - i s') ] ) e^{-(c+i s') \ln(x)} ds.
\end{equation}
So (\ref{integral}) can also be evaluated as an inverse Fourier transform (or inverse two-sided Laplace transform):
$$ f_X(x) = \frac{e^{-c \ln(x/\sigma r^{\frac{1}{\alpha}}) }}{\alpha \pi \sigma r^{\frac{1}{\alpha}}} \mathcal{F}_{\ln(x/\sigma r^{\frac{1}{\alpha}})}^{-1} \left[  \G{c+ is' } \G{ \frac{1- c - i s' }{\alpha} } \sin( \pi [ \gamma - \gamma (c - i s' )] ) \right].$$

There are several well-known ways of inverting Fourier and Laplace transforms numerically (see e.g.\ \cite{kuznetsov2013convergence} and references therein). The naive approach is simply to truncate the contour at the points $ \pm b$ (or $c \pm i b$ in the original coordinates). This truncated approximation is indeed practical here, yielding
\begin{equation}\label{Cgamma_bound}
|\Gamma(a+ib)|^{2}=|\Gamma(a)|^{2}\prod _{k=0}^{\infty }{\frac {1}{1+{\frac {b^{2}}{(a+k)^{2}}}}}
\end{equation}

That is, as the imaginary part of the contour increases in magnitude, the complex Gamma function decays rapidly. 

The reader can implement these inversions using their preferred numerical methods. Additionally, there are commercial numerical Mellin/Fourier/Laplace inversion packages available for example in computing systems such as Mathematica and Matlab 

\paragraph{Series Representations} 

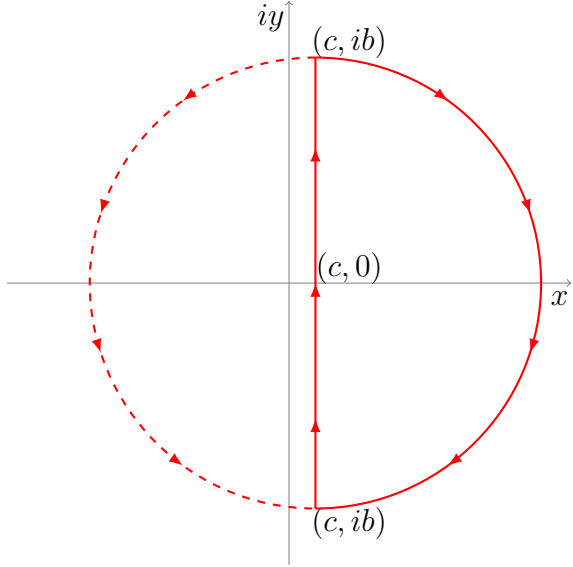
\begin{figure}
\begin{center}
\begin{tikzpicture}
\def\gap{10}
\def\bigradius{3}

\draw [help lines,->] (-1.25*\bigradius, 0) -- (1.25*\bigradius,0);
\draw [help lines,->] (0, -1.25*\bigradius) -- (0, 1.25*\bigradius);

\node at (3.6,-0.2){$x$};
\node at (-0.24,3.53) {$iy$};

\node at (0.8,0.2) {$(c,0)$}; 
\node at (0.8,\bigradius + 0.2) {$(c,ib)$}; 
\node at (0.8,-\bigradius - 0.2) {$(c,ib)$}; 

\draw[thick,red,xshift=\gap,
decoration={ markings,  
      mark=at position 0.2 with {\arrow{latex}}, 
      mark=at position 0.5 with {\arrow{latex}},
      mark=at position 0.8 with {\arrow{latex}}}, 
      postaction={decorate}]
  (0,-\bigradius)  -- (0,\bigradius);

\draw[thick,red,xshift=\gap,
decoration={ markings,
      mark=at position 0.2 with {\arrow{latex}}, 
      mark=at position 0.4 with {\arrow{latex}},
      mark=at position 0.6 with {\arrow{latex}}, 
      mark=at position 0.8 with {\arrow{latex}}}, 
      postaction={decorate}]
 (0,\bigradius) arc (90:-90:\bigradius) -- (0,-\bigradius);
  
\draw[thick, dashed, red ,xshift=\gap,
decoration={ markings,
      mark=at position 0.2 with {\arrow{latex}}, 
      mark=at position 0.4 with {\arrow{latex}},
      mark=at position 0.6 with {\arrow{latex}}, 
      mark=at position 0.8 with {\arrow{latex}}}, 
      postaction={decorate}]
 (0,\bigradius) arc (-90:90:-\bigradius) -- (0,-\bigradius);

\end{tikzpicture}
\end{center}
\caption{Possible contours}
\label{Circles}
\end{figure}

Deriving power series and asymptotic expansions directly from (\ref{integral}) is a relatively simple application of complex analysis. Consider the two half-circles in the complex plane formed from the diameter running between $c \pm i b$ where $c \in (0,1)$ and $b\in \mathbb{R}$ (Fig.\ \ref{Circles}).

As we allow $b \rightarrow \infty$, the segment of the contour along the diameter will become what we need to evaluate (\ref{integral}). Provided the contribution from the arc of the half-circle disappears in the usual way, can use Cauchy's Residue Theorem to evaluate the contour integral. From (\ref{Cgamma_bound}) we know that the modulus of the integrand of (\ref{integral}) on the half circle will achieve its maximum on the point intersecting the real line. The remaining issue is to decide which half-circle and therefore what residues to use. Assume $\alpha>1$ and consider two distinct limits of the Gamma function product:


\noindent\begin{minipage}{.5\linewidth}

\begin{align*}
& \hspace{5 mm} \G{s} \G{ \frac{1-s}{\alpha} } \\
&= \frac{\G{ \frac{1-s}{\alpha} }}{\G{1-s}} \frac{\pi}{\sin(\pi[1-s])} \\
& \xrightarrow[s \rightarrow -\infty]{\alpha>1} 0\\
\end{align*}

\end{minipage}%
\begin{minipage}{.5\linewidth}

\begin{align*}
& \hspace{5 mm}\G{s} \G{ \frac{1-s}{\alpha} }\\
& = \frac{\G{s}}{\G{ 1- 1/\alpha + s/\alpha }} \frac{\pi}{\sin(\pi [\frac{1-s}{\alpha}]) } \\
& \xrightarrow[s \rightarrow \infty]{ \alpha>1} \infty\\
\end{align*}

\end{minipage}

We can see that for the integrand in (\ref{integral}), if we integrate along the left semi-circle, in the limit of an infinitely large radius the contribution from the left arc disappears by the estimation lemma and we recover (\ref{integral}). This allows us via residues to easily derive series representations:

\begin{align*}
&\hspace{ 5mm} \frac{1}{\alpha \pi \sigma r^{\frac{1}{\alpha}}}  \frac{1}{2\pi i} \left(\int_{\text{diameter}} \tilde{f}(s) ds + \int_{\text{arc}} \tilde{f}(s) ds \right) \\
& \rightarrow   \frac{1}{\alpha \pi}\frac{1}{2 \pi i} \int_{c-i \infty}^{c+i\infty}\tilde{f}(s) ds \hspace{2 mm }+\hspace{2 mm } 0 \\ 
&=\sum_{k=0}^\infty \operatorname{Res} (\tilde{f}(s),-k)
\end{align*}
where
$$\tilde{f}(s)=\G{s} \G{ \frac{1-s}{\alpha} } \sin( \pi [ \gamma - \gamma s ] ) \left(\frac{x}{\sigma r^{\frac{1}{\alpha}}} \right)^{-s} $$
and given $\operatorname{Res} (f,-k)=\lim _{z\to c}(z+k)f(z) $, we have
$$ \operatorname{Res} (\tilde{f}(s),-k) = \operatorname{Res} (\G{s},-k) \G{ \frac{1+k}{\alpha} } \sin( \pi [ \gamma + \gamma k ] ) \left(\frac{x}{\sigma r^{\frac{1}{\alpha}}} \right)^{k}.  $$

For any $k \in \mathbb{N}$ we can use the recurrence formula:
$$(z+k)\Gamma (z)=\frac {\Gamma (z+k+1)}{z(z+1)\cdots (z+k-1)}.$$

The numerator at $z=-k$ is $\Gamma (-k+k+1)=\Gamma (1)=1$ and the denominator is $(-1)^{k}k!$. So the residues of the gamma function at those points are
$$\operatorname {Res} (\Gamma ,-k)={\frac {(-1)^{k}}{k!}} $$
which finally gives
\begin{align*}\label{series1}
f_X(x) &= \frac{1}{\alpha \pi \sigma r^{\frac{1}{\alpha}} }H_{{2,2}}^{{\,1,1}}\!\left[\frac{x}{\sigma r^{\frac{1}{\alpha}} }  
\left|{\begin{matrix}
(1-\frac{1}{\alpha} , \frac{1}{\alpha})&(1 -  \gamma,\gamma)\\
(0,1)&(1-  \gamma,\gamma)
\end{matrix}}\right.\right] \\
&=
\frac{1}{\alpha \pi \sigma r^{\frac{1}{\alpha}} } \sum_{k=0}^\infty\frac{\G{ \frac{1+k}{\alpha} } \sin( \pi [ \gamma + \gamma k ] ) }{k!}\left(-\frac{x}{\sigma r^{\frac{1}{\alpha}} } \right)^{k} \text{,   \hspace{5mm}  $\alpha>1$. } 
\end{align*}

For $|x|\leq 1$ this series will converge rather slowly as it will take many terms for the gamma function to overpower the $x^k$ term. This would make some kind of asymptotic series desirable. Interestingly the series for the $\alpha<1$ case is such a series. 

If we instead examine the opposite half circle (with residues $s=1+ \alpha k$) we would derive the series for the case of $\alpha<1$: 
$$ \operatorname{Res} (\tilde{f}(s),1+ \alpha k) = \G{s} \operatorname{Res}\left(\G{ \frac{1+k}{\alpha} }, 1+ \alpha k \right) \sin( \pi [ \gamma - \gamma (1+ \alpha k)  ] ) \left(\frac{x}{\sigma r^{\frac{1}{\alpha}} } \right)^{-1-\alpha k}.  $$

Using the Recurrence Formula in a similar fashion yields 
\begin{equation}\label{series2}
f_X(x)=\frac{1}{\alpha \pi \sigma r^{\frac{1}{\alpha}} } \sum_{k=0}^\infty\frac{\G{ 1+ \alpha k } \sin( \pi [ \gamma + \gamma (1+ \alpha k) ] ) }{k!}(-1)^{k} \left(\frac{x}{\sigma r^{\frac{1}{\alpha}} } \right)^{-1-\alpha k} \text{,   \hspace{5mm}  $\alpha<1$. } 
\end{equation}

To see how this will give an asymptotic series in the $\alpha>1$ case requires the use of a modification of the standard Jordan's Lemma (for proof see Appendix (\ref{sec:Jordan Proof})). 

\begin{lemma}[Jordan's Lemma] 

Given the right-hand semicircle in Fig.\ \ref{Circles} with the part-contour $ arc=\{c+ Re^{-i\theta }\mid \theta \in [-\frac{\pi}{2},\frac{\pi}{2} ] \} $, 

$$\left|\int _{arc}e^{-az}g(z)\,dz\right|\leq e^{-a}{\frac {\pi }{a}}M_{R}\quad {\text{where}}\quad M_{R}:=\max _{\theta \in [-\frac{\pi}{2},\frac{\pi}{2}]}\left|g\left(c+Re^{i\theta }\right)\right|. $$

\end{lemma}

We have from (\ref{integral}) that $a=\ln(x/\sigma r^{\frac{1}{\alpha}})$ and since $g(\cdot)$ acheives it's max on $\theta=0$ $M_R = \G{c+R} \G{ \frac{1-c-R}{\alpha} } \sin( \pi [ \gamma - \gamma(c+R) ] )$. So we have ${\frac {e^{-a}\pi M_{R}}{a}} \rightarrow 0$ as $x\rightarrow \infty$.  Since the contribution from the arc vanishes for large $x$ regardless of $R$, that leaves just the contribution from the diameter, and thus as $x\rightarrow \infty$, 
\begin{equation}\label{series3}
f_X(x)\sim \frac{1}{\alpha \pi \sigma r^{\frac{1}{\alpha}} }\sum_{k=0}^\infty\frac{\G{ 1+ \alpha k } \sin( \pi [ \gamma + \gamma (1+ \alpha k) ] ) }{k!}(-1)^{k} \left(\frac{x}{\sigma r^{\frac{1}{\alpha}} } \right)^{-1-\alpha k} \text{, $\alpha>1$. }
\end{equation}

\paragraph{The TCE Case} Following the same steps as above, we can show for example that
\begin{align*}
&\frac{\sigma r^{\frac{1}{\alpha}} H_{{2,2}}^{{\,1,1}}\!\left[\frac{s_q}{\sigma r^{\frac{1}{\alpha}}}  
\left|{\begin{matrix}
(1-\frac{\alpha-1}{\alpha} , \frac{1}{\alpha})&(\gamma,\gamma)\\
(0,1)&(\gamma,\gamma)
\end{matrix}}\right.\right]}{1- q} \\
&\qquad = \frac{\sigma r^{\frac{1}{\alpha}}}{1- F_S(s)}\sum_{k=0}^\infty \frac{\G{1- \frac{1}{\alpha} + \frac{k}{\alpha} } \sin( \pi [ 1- \gamma + \gamma k  ] ) }{  k!} \left(- \frac{s_q}{\sigma r^{ \frac{1}{\alpha} }} \right)^k .
\end{align*}

An asymptotic series is again available in the $\alpha<1$ case; however, Fig.\ \ref{Circles} no longer applies. Instead, the Bromwich path in the $\alpha<1$ case must separate poles at $s=-k$ and $\alpha - 1 - \alpha k$. 

As a general comment, we would recommend the integral transform inversion approach whenever possible. The main issues are that the transition from the truncated power series to the asymptotic regime can create inaccuracies at points of interest (e.g.\ at high $q$ values in the TCE case). 

\section{Conclusion}\label{SEC: Conclusion}

In this \ifbool{is_paper}{paper}{chapter} we have presented a new way of computing allocations among business units with shared systemic shocks, via the GCLT and some useful properties of Stable distributions. Notably, we make use of the fact that as long as our business units share a tail index we can make a Stable approximation. 

The methods presented here allow for a very general, prudent and simple way of handling the allocation problem. By assuming the worst-case tail index for each loss, the Stable approximation is viable. This in turn leads to very simple allocations as linear functions of total risk capital. Additionally, the total risk capital has been computed in the Stable approximation for the first time in the TCE case using the Fox H-functions. 

The possibility of computing TCE allocation estimates should be valuable for insurance companies and banks trying to define their minimum capital requirements under the Basel framework. This approach may also be generalized for other useful risk functionals.

\newpage

 
\bibliographystyle{apalike}
\bibliography{York_thesis}

\newpage

\appendix

\let\part\section
\let\section\subsection

\part{Univariate Stable Background}\label{app: AppendixA}
\begin{definition}[Stable Random Variable, 1st definition] \label{def: Stable_Def_1}

A random variable $X$ is said to have a \textit{stable distribution} if for $n\ge2$, $\exists c_n\in \mathbb{R}^+, d_n \in \mathbb{R} $ such that:

\begin{equation}\label{eq: stable_eq}
X_1 + ... + X_n \stackrel{d}{=} c_n X + d_n
\end{equation}

where the $X_i$ are independent copies of $X$.\\

\end{definition}

\noindent One can derive from this definition (\cite{zolotarev1986one}) that the characteristic function for a stable variable $X$ is given by:
\begin{equation}\label{Stable_CF}
	\phi_{X}(t) = 
	\begin{cases}
		\exp(-|\sigma t|^\alpha ( 1 - i \beta (\sign t) a ) + i t \mu ) & \alpha \ne 1 , \\
		\exp(-|\sigma t| ( 1 + i \beta \frac{2}{\pi}(\sign t) \ln |t| ) + i t \mu ) & \alpha=1 . \\
	\end{cases}
\end{equation}
Denoted $X \sim S_\alpha(\mu,\sigma,\beta)$  with $a=\tan (\frac{\pi \alpha}{2} )$. The $\mu$ and $\sigma$ are the location and scale parameters, which are equal and proportional to the mean and variance, respectively, whenever they exist. Here $\alpha \in (0,2]$ is the tail parameter. For values of $\alpha=2$ we have a Normal distribution, and for $\alpha<2$ a Pareto tailed distribution with exponent $\alpha$. The value $\beta$ is a skewness parameter and if $\alpha<1$ and $\beta = \pm 1$ support is either $[\mu,\infty)$ or $(-\infty, \mu]$. Finally, Stable random variables behave under addition operation much like the Normal. For $X_1 \sim S_\alpha(\mu_1,\sigma_1,\beta_1)$ and $X_2 \sim S_\alpha(\mu_2,\sigma_2,\beta_2)$, then $X_1 + X_2 \sim S_\alpha(\mu,\sigma,\beta)$ where:
\begin{align} 
\label{sum_prop}
\begin{split}
\mu &=\mu _{1}+\mu _{2}, \\
\sigma &=\left(\sigma_{1}^{\alpha }+\sigma_{2}^{\alpha }\right)^{\frac {1}{\alpha }}, \\
\beta &=\frac {\beta _{1}\sigma_{1}^{\alpha }+\beta _{2}\sigma_{2}^{\alpha }}{\left(\sigma_{1}^{\alpha }+\sigma_{2}^{\alpha }\right)}.
\end{split}
\end{align}

First appearing in Paul L\'{e}vy's 1925 monograph \textit{Calcul des probabilit\'{e}s}, Stable distributions went on to be studied by leading researchers, such as Andrey Kolmogorov and William Feller. One of the motivating problems of probability theory has been the distribution of sums of random variables. Stable random variables generalize the Normal as a basin of attraction to encompass all i.i.d. sums, not just the ``nice" ones with finite variance or bounded support:

\begin{theorem}{ (The generalized central limit theorem) }\label{thm: GCLT} Consider the sequence of centred and normalized sums of i.i.d RVs $ Y_{i}$ with Pareto tails such that:

\begin{center}
$1-F_{Y_i}(y) \sim k_1 y^{-\alpha}$ and $F_{Y_i}(y) \sim k_2 |y|^{-\alpha}$
\end{center}

Define: 

$$ Z_n = \frac{Y_1 + ... + Y_n }{p_n} - q_n $$

and for $\alpha \neq 1,2$\footnote{In the normal case $B_n= \sqrt{n}$. See \cite{uchaikin2011chance} for Cauchy case} set:

\begin{center}
$p_n^\alpha = \frac{2 \G{\alpha} sin(\alpha \pi /2) }{ \pi (C_1+C_2)}n$ and $q_n=\expect[Y_i]$ (if it exists, zero otherwise)
\end{center}

Then $f_{Z_n} \rightarrow f_S$ weakly where $f_S$ is a standardized stable distribution. i.e

$$ Z_n \xrightarrow[]{dist.} S_\alpha(1,\beta,0)$$

\end{theorem}

While extremely useful, Stable distributions have historically been less popular than other models. This is likely due to the fact that Stable PDFs generally do not exist in closed form. There are however three notable cases where this is not true:
\begin{itemize}
\item Normal $(\alpha=2)$;

\item Cauchy (t with d.o.f=1) $(\alpha=1, \beta=0)$;

\item L\'{e}vy $(\alpha = \frac{1}{2}, \beta=1)$.
\end{itemize}

\part{H-Fun Background}\label{app: AppendixB}

\section{The Fox H-function }\label{SEC:H_ref}

Here we give a brief introduction to H-functions. For more details, refer to \cite{springer1979algebra}. 

\begin{definition}[The Fox H-function]\label{def:H Fun}
For $0 \leq m \leq q$, $0 \leq n \leq p$ and $A_j, B_j >0$,
\begin{align*}
& \hspace{0.5 cm} H_{{p,q}}^{{\,m,n}}\!\left[z\left|{\begin{matrix}(a_{1},A_{1})&(a_{2},A_{2})&\ldots &(a_{p},A_{p})\\(b_{1},B_{1})&(b_{2},B_{2})&\ldots &(b_{q},B_{q})\end{matrix}}\right.\right] \\
&\qquad\qquad =\mathcal{M}_{z}^{-1} \left\lbrace {\frac  {(\prod _{{j=1}}^{m}\Gamma (b_{j}+B_{j}s))(\prod _{{j=1}}^{n}\Gamma (1-a_{j}-A_{j}s))}{(\prod _{{j=m+1}}^{q}\Gamma (1-b_{j}-B_{j}s))(\prod _{{j=n+1}}^{p}\Gamma (a_{j}+A_{j}s))}} \right\rbrace \\
&\qquad\qquad ={\frac  {1}{2\pi i}}\int _{C}{\frac  {(\prod _{{j=1}}^{m}\Gamma (b_{j}+B_{j}s))(\prod _{{j=1}}^{n}\Gamma (1-a_{j}-A_{j}s))}{(\prod _{{j=m+1}}^{q}\Gamma (1-b_{j}-B_{j}s))(\prod _{{j=n+1}}^{p}\Gamma (a_{j}+A_{j}s))}}z^{{-s}}\,ds
\end{align*}
where $C$ separates the poles of products of Gamma functions in the numerator such that it lies respectively to the right and left of
\begin{center}
$z = \dfrac{b_j +k}{B_j}$  \hspace{10 mm}  $z= \dfrac{a_j -1 - k}{A_j}$
\end{center}
for all $j$ and $k=1,2,3...$. 

For $z\in\mathbb{R}$ the function is only defined for $z>0$. 

\end{definition}

\paragraph{Important Properties:}
\begin{enumerate}
\item For $c>0$,
$$\hspace{0.5 cm} H_{{p,q}}^{{\,m,n}}\!\left[z^c\left|{\begin{matrix}(a_{1},A_{1}) &\ldots &(a_{p},A_{p})\\(b_{1},B_{1})&\ldots &(b_{q},B_{q})\end{matrix}}\right.\right] \\
=\frac{1}{c}H_{{p,q}}^{{\,m,n}}\!\left[z\left|{\begin{matrix}(a_{1},\dfrac{A_{1}}{c}) &\ldots &(a_{p},\dfrac{A_{p}}{c})\\  \\(b_{1},\dfrac{B_{1}}{c})&\ldots &(b_{q},\dfrac{B_{q}}{c})\end{matrix}}\right.\right]. $$

For $c<0$, use the fact that
$$H_{{p,q}}^{{\,m,n}}\!\left[\frac{1}{z}\left|{\begin{matrix}(a_{1},A_{1}) &\ldots &(a_{p},A_{p})\\(b_{1},B_{1})&\ldots &(b_{q},B_{q})\end{matrix}}\right.\right] \\
= H_{{q,p}}^{{\,n,m}}\!\left[z\left|{\begin{matrix}(1-b_{1},B_{1})&\ldots &(1-b_{q},B_{q})\\(1-a_{1},A_{1}) &\ldots &(1-a_{p},A_{p})\end{matrix}}\right.\right]. $$

\item For $d\in \mathbb{R}$, 
$$z^d H_{{p,q}}^{{\,m,n}}\!\left[z\left|{\begin{matrix}(a_{1},A_{1}) &\ldots &(a_{p},A_{p})\\(b_{1},B_{1})&\ldots &(b_{q},B_{q})\end{matrix}}\right.\right] \\
=H_{{p,q}}^{{\,m,n}}\!\left[z\left|{\begin{matrix}(a_{1}+dA_1,A_{1}) &\ldots &(a_{p}+d A_p,A_{p})\\(b_{1}+d B_1,B_{1})&\ldots &(b_{q}+d B_1,B_{q})\end{matrix}}\right.\right]. $$

\item The Laplace Transform is also an H-function:
\begin{align*}
&\lap{ r }{ H_{{p,q}}^{{\,m,n}}\!\left[cz\left|{\begin{matrix}(a_{1},A_{1}) &\ldots &(a_{p},A_{p})\\(b_{1},B_{1})&\ldots &(b_{q},B_{q})\end{matrix}}\right.\right]} \\
&\qquad\qquad=\frac{1}{c}H_{{p,q}}^{{\,m,n}}\!\left[\frac{r}{c}\left|{\begin{matrix}(1-b_{1}-B_1,B_{1})&\ldots &(1-b_{q}-B_q,B_{q})\\(1-a_{1}-A_1,A_{1}) &\ldots &(1-a_{p}-A_p,A_{p})\end{matrix}}\right.\right].
\end{align*}

\end{enumerate}

\newpage
\section{H-Function Representation of One-Dimensional Stable Densities}\label{sec:pdf_proof}

Originally shown by \cite{schneider1986stable} (using a different parameterization of stable distributions) we rederive the Fox H-function representation of a univariate stable distribution in our notation.


\begin{theorem}
For $X \sim S_\alpha(\sigma,\beta,0)$  , $x\geq 0$ , $r=\sqrt{1+(a \beta)^2}$, $\phi =\tan^{-1}(a \beta)  $ and $\gamma = \frac{1}{2} - \frac{\phi}{ \alpha \pi }$,
$$ f_X(x)= \frac{1}{\alpha \pi \sigma r^{\frac{1}{\alpha}} }H_{{2,2}}^{{\,1,1}}\!\left[\frac{x}{\sigma r^{\frac{1}{\alpha}} }  
\left|{\begin{matrix}
\left(1-\dfrac{1}{\alpha} , \dfrac{1}{\alpha}\right)&(1 -  \gamma,\gamma)\\
(0,1)&(1-  \gamma,\gamma)
\end{matrix}}\right.\right] $$

\end{theorem}

\begin{proof}

Let $\xi_t = 1 - i \beta (\sign t) a$, $\xi = 1 - i \beta a$ and $\bar{\xi} = 1 + i \beta a$. We begin by inverting the characteristic function:
\begin{align*}
f(x) &= \frac{1}{2\pi} \int_{- \infty}^\infty e^{-i x t}  e^{ |t|^\alpha \xi_t}   dt  \\
&=\frac{1}{2\pi} \int_{0}^\infty e^{-i x t}  e^{ |t|^\alpha \xi}   dt  +\frac{1}{2\pi} \int_{-\infty}^0 e^{-i x t}  e^{ |t|^\alpha \bar{\xi}}   dt  \\
& =\operatorname {Re}\left[ \frac{1}{\pi} \int_{0}^\infty e^{-i x t}  e^{ t^\alpha \xi}   dt \right] \\
&=\frac{1}{\pi}\operatorname {Re} \left[ \lap{ix}{ e^{- t^\alpha \xi} } \right] \\
&=\frac{1}{\pi}\operatorname {Re} \left[ \frac{1}{ \alpha \xi^\frac{1}{\alpha} } 
H_{{1,1}}^{{\,1,1}}\!\left[ \frac{ix}{\xi^\frac{1}{\alpha}}  \left|{\begin{matrix} \left( 1-\dfrac{1}{\alpha},\dfrac{1}{\alpha}\right) \\(0,1)\end{matrix}}\right.\right]  
\right] \\
& \hspace{0.5 cm} \text{\normalfont by Lemma \ref{Lemma 1}} \\
&=\frac{1}{\alpha \pi \sigma r^{\frac{1}{\alpha}} }H_{{2,2}}^{{\,1,1}}\!\left[\frac{x}{\sigma r^{\frac{1}{\alpha}} }  
\left|{\begin{matrix}
\left(1-\dfrac{1}{\alpha} , \dfrac{1}{\alpha}\right)&(1 -  \gamma,\gamma)\\
(0,1)&(1-  \gamma,\gamma)
\end{matrix}}\right.\right] \\
& \hspace{0.5 cm} \text{\normalfont by Lemma \ref{Lemma 2}, where } \theta_1=\frac{\phi}{\alpha}, \nu_2=\nu_1=-1, \theta_2 = \left( \frac{\pi}{2}-\frac{\phi}{\alpha} \right).
\end{align*}
\end{proof}

\part{Proofs of Lemmas}\label{app: AppendixC}
\section{Proof of Lemma \ref{Lemma 1}}\label{sec:Lem1_proofs}

\begin{proof}

First note that being an exponential function, the \glssymbol{mellin}[Mellin transform] of $e^{-bt^\alpha}$ is a Gamma function:
$$\mel{s}{e^{-bt^\alpha}} \\
= \int_{0}^{\infty} t^ {s-1} e^{-bt^\alpha} dt.$$

Making the substitution $u= bt^\alpha$, we obtain $\frac{dt}{t}=\frac{du}{\alpha u}$, so that
\begin{align*}
\mel{s}{e^{-bt^\alpha}} &= \int_{0}^{\infty} \left( \frac{u}{b} \right)^{s/\alpha} e^{-u} \frac{du}{\alpha u} \\
&= \frac{1}{\alpha b^{\frac{s}{\alpha}} } \int_{0}^{\infty} u^{s/\alpha - 1 } e^{-u} du \\
&= \frac{1}{\alpha b^{\frac{s}{\alpha}} } \Gamma \left( \frac{s}{\alpha} \right). \\ 
\end{align*}
\vspace{-10mm}

Making use of the Mellin-Barnes Representation,
$$e^{-bt^\alpha} = \frac{1}{\alpha} \mathcal{M}^{-1} \left[ \Gamma \left( \frac{s}{\alpha} \right) b^{-\frac{s}{\alpha}} \right] = \frac{1}{\alpha} H_{{0,1}}^{{\,1,0}}\!\left[b^\frac{1}{\alpha} t \left|{\begin{matrix} - \\ \left( 0,\frac{1}{\alpha}\right) \end{matrix}}\right.\right].$$

Using Property 2 of Fox H-functions from Appendix \ref{SEC:H_ref},
\begin{align*}
t^j e^{-bt^\alpha} &= \frac{t^j}{\alpha} H_{{0,1}}^{{\,1,0}}\!\left[b^\frac{1}{\alpha} t \left|{\begin{matrix} - \\ \left( 0,\dfrac{1}{\alpha}\right) \end{matrix}}\right.\right] \\
&= \frac{ (b^\frac{1}{\alpha} t)^j }{\alpha b^\frac{j}{\alpha}} H_{{0,1}}^{{\,1,0}}\!\left[b^\frac{1}{\alpha} t \left|{\begin{matrix} - \\ \left( 0,\frac{1}{\alpha}\right) \end{matrix}}\right.\right] \\
&= \frac{1}{\alpha b^\frac{j}{\alpha}} H_{{0,1}}^{{\,1,0}}\!\left[b^\frac{1}{\alpha} t \left|{\begin{matrix} - \\ \left( \frac{j}{\alpha},\frac{1}{\alpha}\right) \end{matrix}}\right.\right]. \\
\end{align*}
\vspace{-10mm}

Finally, using the Laplace transform (Property 3 in Appendix \ref{SEC:H_ref}) yields
$$\lap{x}{ \frac{1}{\alpha b^\frac{j}{\alpha}} H_{{0,1}}^{{\,1,0}}\!\left[b^\frac{1}{\alpha} t \left|{\begin{matrix} - \\ \left( \frac{j}{\alpha},\frac{1}{\alpha}\right) \end{matrix}}\right.\right] } 
=\frac{1}{\alpha b^\frac{j+1}{\alpha}} 
H_{{1,1}}^{{\,1,1}}\!\left[ \frac{x}{b^\frac{1}{\alpha}} \left|{\begin{matrix} \left( 1-\dfrac{j+1}{\alpha},\dfrac{1}{\alpha}\right) \\(0,1)\end{matrix}}\right.
\right].$$
\end{proof}

 \section{Proof of Lemma \ref{Lemma 2}}\label{sec:Lem2_proofs}

\begin{proof}

Define
\begin{align*}
z_1^{\nu_1} &= r_1^{\nu} e^{i \theta_1 \nu_1} \\
z_2^{ \nu_2 } &= r_2^{ \nu_2 } e^{i \theta_2 \nu_2}
\end{align*}
and recall the Euler reflection formula:
$$ \Gamma (1-z)\Gamma (z)={\pi  \over \sin(\pi z)}.$$ 

We have
\begin{align*}
& z_1^{\nu_1}
H_{{1,1}}^{{\,1,1}}\!\left[ z_2^{\nu_2} x  \left|{\begin{matrix} \left( a_1,A_1\right) \\(b_1,B_1)\end{matrix}}\right.\right]  
 \\
&\quad={\frac  {z_1^{\nu_1}}{2\pi i}}\int _{C} \Gamma (b_1+B_1 s) \Gamma (1-a_1-A_1 s))(z_2^{ \nu_2 } x)^{{-s}} \,ds \\
&\quad={\frac  {1}{2\pi i}}\int _{C} \Gamma (b_1+B_1 s) \Gamma (1-a_1-A_1 s)) \left( r_1^{\nu_1} e^{ i( \theta_1 \nu_1  - \theta_2 \nu_2  s ) }  \right) (r_2^{ \nu_2 } x)^{{-s}} \,ds \\
&\quad= {\frac  { r_1^{\nu_1}}{2\pi i}}\int _{C} \Gamma (b_1+B_1 s) \Gamma (1-a_1-A_1 s))   \sin\left( \pi \left(\frac{1}{2} +  \frac{\theta_1 \nu_1}{\pi} - \frac{\theta_2 \nu_2}{\pi}s  \right) \right)  (r_2^{ \nu_2 } x)^{{-s}} \,ds \\
&\quad\quad + i {\frac  { r_1^{\nu_1}}{2\pi i}}\int _{C} \Gamma (b_1+B_1 s) \Gamma (1-a_1-A_1 s))   \sin\left( \pi \left( \frac{\theta_1 \nu_1}{\pi} - \frac{\theta_2 \nu_2}{\pi}s   \right) \right)   (r_2^{ \nu_2 } x)^{{-s}} \,ds \\
& \text{and using the definition of the Fox H-function and the reflection formula,} \\
&\quad= \pi r_1^{\nu_1} H_{{2,2}}^{{\,1,1}}\!\left[ r_2^{\nu_2} x  \left|{\begin{matrix} \left( a_1,A_1\right) & \left( \frac{1}{2}-\frac{\theta_1 \nu_1}{\pi},{\frac{\theta_2 \nu_2}{\pi}}\right)  \\ (b_1,B_1) &  \left( \frac{1}{2}-\frac{\theta_1 \nu_1}{\pi},{\frac{\theta_2 \nu_2}{\pi}}\right)  \end{matrix}}\right.\right]  
 \\
&\quad\quad+ i \pi r_1^{\nu_1} H_{{2,2}}^{{\,1,1}}\!\left[ r_2^{\nu_2} x  \left|{\begin{matrix} \left( a_1,A_1\right) & \left(1- \frac{\theta_1 \nu_1}{\pi},{\frac{\theta_2 \nu_2}{\pi}}\right)  \\ (b_1,B_1) &  \left(1- \frac{\theta_1 \nu_1}{\pi},{\frac{\theta_2 \nu_2}{\pi}}\right)  \end{matrix}}\right.\right]  
 \\
\end{align*}
\end{proof}

\section{Proof of Modified Jordan's Lemma in Section \ref{SEC:Evaluation}}\label{sec:Jordan Proof}

\begin{proof}
Consider the CW curve $\Upsilon: s=c+Re^{-i\theta}, \hspace{3 mm} \theta \in [-\frac{\pi}{2},\frac{\pi}{2}]$.  
\begin{align*}
&\left\vert \int_\Upsilon g(s) z^{-s} ds \right\vert \\
&\quad= \left\vert \int_\Upsilon g(s) e^{-s\ln(z)} ds \right\vert \\
&\quad= e^{-c\ln(z)} \left\vert \int_{-\frac{\pi}{2}}^{\frac{\pi}{2}} g(c+Re^{-i\theta}) \exp{ \{ -\ln(z)[ R\cos(\theta) - i R \sin(\theta)] \} } (-iR) e^{-i\theta}  d\theta \right\vert \\
&\quad\leq R M_R e^{-c\ln(z)} \int_{-\frac{\pi}{2}}^{\frac{\pi}{2}} e^{ -\ln(z) R\cos(\theta)  } d\theta \\
&\quad= 2 R M_R e^{-c\ln(z)} \int_{-\frac{\pi}{2}}^{0} e^{ -\ln(z) R\cos(\theta)  } d\theta \\
&\quad\leq 2 R M_R e^{-(c+R)\ln(z)} \int_{-\frac{\pi}{2}}^{0} e^{ -2R\ln(z) \theta/\pi  } d\theta \\
&\qquad\qquad \text{(since on $[-\pi/2,0]$ we have $-\cos(\theta) < -1-2\theta/\pi$)  } \\
&\quad= M_R e^{-(c+R)\ln(z)} \left( \frac{\pi}{\ln(z)} \left( e^{R\ln(z)} -1 \right) \right) \\
&\quad\leq M_R \frac{\pi}{\ln(z)} e^{-c\ln(z)}.
\end{align*}

\end{proof}

\end{document}